\documentclass[twocolumn,aps,showpacs,amsmath,amssymb]{revtex4}

\usepackage{graphicx}
\usepackage{dcolumn}
\usepackage{bm}

\begin{document}

\title{
Improved parametrization of the unified model for $\alpha$ decay and $\alpha$ capture}%

\author{
V. Yu. Denisov$^{1,2}$, O. I. Davidovskaya$^{1}$, I. Yu. Sedykh$^{3}$
}%

\affiliation{%
$^{1}$ Institute for Nuclear Research, Prospect Nauki 47,
03680 Kiev, Ukraine \\
$^{2}$ Faculty of Physics, Taras Shevchenko National University of Kiev, Prospect Glushkova 2, 03022 Kiev, Ukraine\\
$^{3}$ Department of Mathematics, Financial University, Leningradsky Prospect 49, 125993 Moscow, Russia
}%

\date{\today}

\begin{abstract}
The updated data for the ground-state-to-ground-state $\alpha$-transition half-lives in 401 nuclei and the $\alpha$ capture cross sections of $^{40}$Ca, $^{44}$Ca, $^{59}$Co, $^{208}$Pb and $^{209}$Bi are well described in the framework of the unified model for $\alpha$-decay and $\alpha$-capture. The updated values of the $\alpha$ decay half-lives, the binding energies of nuclei, the spins of parent and daughter nuclei, and the surface deformation parameters are used for the reevaluation of the model parameters. The data for the ground-state-to-ground-state $\alpha$-decay half-lives are also well described by the empirical relationships.
\end{abstract}

\pacs{24.10.-i, 25.70.-z, 25.70.Jj}

\maketitle

\section{Introduction}

$\alpha$-decay is very important process in nuclear physics \cite{gamov,cg,gn,vs,ame2012,nudat,back,mnk,royer,di,gupta,brasil2,dasgupta,csb,delion,dk-adndt,dk-gs-es,dk-she,dk-emp,pgg,dp,sb,ppg,royer2010,royer2011,ia,ren,santhosh,dd,wnlg}. The theory of $\alpha$-decay was formulated by Gamow \cite{gamov} and independently by Condon and Gurney \cite{cg}. Subsequently various phenomenological and microscopic approaches to the description of $\alpha$-decay were proposed; see, for example, Refs. \cite{gamov,cg,gn,vs,ame2012,nudat,back,mnk,royer,di,gupta,brasil2,dasgupta,csb,delion,dk-adndt,dk-gs-es,dk-she,dk-emp,pgg,dp,sb,ppg,royer2010,royer2011,ia,ren,santhosh,dd,wnlg} and papers cited therein. The simple empirical relationships for description of $\alpha$-decay half-lives have been discussed too, see, for example, Refs. \cite{gn,vs,mnk,dasgupta,royer,brasil2,gupta,dk-emp,pgg,royer2010,royer2011,wnlg} and numerous references therein.

The $\alpha$-decay process involves sub-barrier penetration of $\alpha$-particles through the barrier, caused by interaction between $\alpha$-particle and nucleus. The fusion ($\alpha$-capture) reaction between the $\alpha$-particle and nucleus proceeds in the opposite direction to the $\alpha$-decay reaction. However, the same $\alpha$-nucleus interaction potential is the principal ingredient to describe both reactions. Therefore it is natural to use data for both the $\alpha$-decay half-lives and the near barrier $\alpha$-capture reactions for determination of the $\alpha$-nucleus interaction potential. The $\alpha$-decay half-lives and the $\alpha$-capture cross sections are evaluated in the framework of the unified model for $\alpha$-decay and $\alpha$-capture (UMADAC) in Refs. \cite{di,dk-adndt,dk-gs-es,dk-she}.

The experimental information on the $\alpha$-decay is extensive. The experimental data are continually being expanded and refined. The data have been updated very recently \cite{nudat}. The ground-state masses and spins of nuclei have been updated continually too \cite{ame2012,wallet}. More accurate atomic mass values lead to more exact definition of the $Q$ values of the ground-state-to-ground-state $\alpha$ transitions. More precise definition of the angular momentum value of the emitted $\alpha$ particle relates to accurate values of the ground-state spins. Therefore, it is reasonable to use extended and updated the data for ground-state-to-ground-state $\alpha$-decay transitions \cite{nudat,dk-adndt}, atomic masses, and spins \cite{ame2012,wallet} for more accurate and full description of the $\alpha$-decay half-lives in the framework of UMADAC. The data for the $\alpha$-decay half-lives for 401 nuclei and the $\alpha$-capture cross sections of $^{40}$Ca, $^{44}$Ca, $^{59}$Co, $^{208}$Pb, and $^{209}$Bi \cite{subfus-exp-ca1,subfus-exp-ca2,subfus-exp-co,subfus-exp-pb} are used now. In comparison to this, our previous dataset \cite{dk-adndt,dk-emp} contained 344 cases for the ground-state-to-ground-state $\alpha$-transitions.

It is very important to have simple and accurate expressions for evaluation of $\alpha$-decay half-lives which can be used very easily. The first empirical formula for $\alpha$-decay half-lives was presented by Geiger and Nuttall \cite{gn}. Since then, many other empirical relationships have been proposed by various authors; see, for example, Refs. \cite{vs,royer,gupta,brasil2,dasgupta,dk-emp,royer2010,royer2011,wnlg} and papers cited therein.

In general, experimentalists would like to evaluate expected half-life values during experiment design. This is especially important for $\alpha$-decay studies of superheavy elements \cite{hofmann,oganessian,gregorich,morita} or nuclei, which are very far from the stability line \cite{andreev,kalaninova}, because such processes are very rare and difficult to observe. That is why, simple and accurate empirical relationships are still claimed. As a consequence, many empirical expressions have appeared during the last several decades \cite{royer,gupta,brasil2,dasgupta,dk-emp,pgg,royer2010,royer2011,wnlg}. The new extended data for the $\alpha$-decay half-lives, atomic masses, and spins give the possibility of improving the accuracy of the empirical relationships introduced in Ref. \cite{dk-emp}.

The main points of UMADAC is briefly presented in Sec. II. A discussion of the input data, adjustable parameters and results is given in Sec. III. Secion IV is dedicated to consideration of the simple empirical relationships for description of the $\alpha$-decay half-lives. The conclusion is given in Sec. V.

\section{UMADAC}

The $\alpha$-decay half-life $T_{1/2}$ in UMADAC depends on the total width $\Gamma$ of the $\alpha$-emission \cite{di,dk-adndt,dk-gs-es,dk-she}
\begin{eqnarray}
T_{1/2} = \hbar \ln(2)/\Gamma,
\end{eqnarray}
where
\begin{eqnarray}
\Gamma = \frac{1}{4\pi} \int \gamma(\theta,\phi) d\Omega
\end{eqnarray}
is the total width of decay, $\gamma(\theta,\phi)$ is the partial width of the $\alpha$ emission in the direction $\theta$ and $\phi$, and $\Omega$ is the space angle.
We consider the axial-symmetric nuclei therefore
\begin{eqnarray}
\gamma(\theta) = \hbar \; 10^\nu \; t(Q,\theta,\ell),
\end{eqnarray}
where $10^{\nu}$ is the $\alpha$-particle frequency of assaults on the barrier, which takes into account $\alpha$-particle preformation; $t(Q,\theta,\ell)$ is the transmission coefficient, which shows the probability of penetration through the barrier; and $Q$ is the released energy at $\alpha$ decay. The transmission coefficient can be obtained in the semiclassical WKB approximation as
\begin{eqnarray}
\lefteqn{t(Q,\theta,\ell) = 1/\{ 1} \\ &+& \left. \exp\left[\frac{2}{\hbar}
\int_{a(\theta)}^{b(\theta)} dr \sqrt{2\mu
\left(v(r,\theta,\ell,Q)-Q\right)} \right] \right\}, \nonumber
\end{eqnarray}
where $a(\theta)$ and $b(\theta)$ are the inner and outer turning points determined from the equations $v(r,\theta,\ell,Q)|_{r=a(\theta),b(\theta)}=Q$, and $\mu$ is the reduced mass. The $\alpha$-nucleus potential $v(r,\theta,\ell,Q)$ consists of Coulomb $v_\mathrm{C}(r,\theta)$, nuclear $v_\mathrm{N}(r,\theta,Q)$ and centrifugal $v_\ell(r)$ parts
\begin{eqnarray}
v(r,\theta,\ell,Q)=v_\mathrm{C}(r,\theta) + v_\mathrm{N}(r,\theta,Q) + v_\mathrm{\ell}(r),
\end{eqnarray}
where
\begin{equation}
v_\mathrm{C}(r,\theta) = \frac{2 Z e^2}{r} \left[1 + \frac{3R^2}{5r^2}\beta_{2} Y_{20}(\theta) \right.
+\left. \frac{3R^4}{9r^4}\beta_{4} Y_{40}(\theta) \right]
\end{equation}
for $r \ge r_\mathrm{m}(\theta)$,
\begin{eqnarray}
v_\mathrm{C}(r,\theta)\approx\frac{2 Z
e^2}{r_\mathrm{m}(\theta)}\left[\frac{3}{2}- \frac{r^2}{2r_\mathrm{m}(\theta)^2}\right.+\frac{3R^2}{5r_\mathrm{m}(\theta)^2}
\beta_{2} Y_{20}(\theta)\times \nonumber\\
\times \left(2 - \frac{r^3}{r_\mathrm{m}(\theta)^3} \right) +\left. \frac{3R^4}{9r_\mathrm{m}(\theta)^4}\beta_{4}
Y_{40}(\theta)\left(\frac{7}{2}-\frac{5r^2}{2r_\mathrm{m}(\theta)^2}\right) \right]
\end{eqnarray}
for $r \le r_\mathrm{m}(\theta)$,
\begin{equation}
v_\mathrm{N}(r,\theta,Q) = \frac{V(Q)}{1+\exp[(r-r_\mathrm{m}(\theta))/d]},
\end{equation}
\begin{equation}
v_\mathrm{\ell}(r) =\frac{\hbar^2 \ell (\ell+1)}{2\mu r^2}.
\end{equation}
Here $Z$, $R$, $\beta_{2}$ and $\beta_{4}$ are, respectively, the number of protons, the radius, the quadrupole and hexadecapole deformation parameters of the nucleus interacting with the $\alpha$-particle; $e$ is the charge of proton, $Y_{20}(\theta)$ and $Y_{40}(\theta)$ are the harmonic functions; $V(Q)$ and $r_\mathrm{m}(\theta)$ are, correspondingly, the strength and the effective radius of the nuclear part of the $\alpha$-nucleus potential. Presentation of the Coulomb field in the form (7) at distances $r \lesssim r_\mathrm{m}(\theta)$ ensures the continuity of the Coulomb field and its derivative at matching point $r=r_\mathrm{m}(\theta)$. The expressions for $V(Q)$, $r_\mathrm{m}(\theta)$ and $d$ are given below.

The angle and angular momentum are the canonical variables in quantum mechanics and do not commute. Therefore a simultaneous specification of the angle and angular momentum is restricted by the uncertainty principle \cite{schiff}. Note the angular momentum of $\alpha$-transition $\ell$ is the precisely specified variable in our approach for $\alpha$-decay. In contrast to this, the angle of $\alpha$-particle emission from a deformed nucleus is not experimentally observed, and we take the averaging on all possible directions of $\alpha$-emission, see Eq. (2). Therefore the simultaneous use of the angle and angular momentum in Eq. (5) with the consequent averaging on the angle of $\alpha$-particle emission may be applicable at the semiclassical evaluation of $\alpha$-decay half-lives. Nevertheless, the canonical variables are treated more consistently in the framework of the exact coupled channels approaches for $\alpha$-decay of deformed nuclei, see, for example, Refs. \cite{delion,sb,dd,ren} and papers cited therein.

The $\alpha$-capture cross section of an axial-symmetric nucleus at near-barrier collision energy $E$ in the center-of-mass system is equal to \cite{di,dk-adndt}
\begin{eqnarray}
\sigma(E)=\frac{\pi \hbar^2}{2\mu E} \int_0^{\pi/2} \sum_\ell
(2\ell+1) t(E,\theta,\ell) \sin(\theta) d\theta.
\end{eqnarray}
Here the integration over angle $\theta$ is done for the same reason as in Eq. (2). The transmission coefficient $t(E,\theta,\ell)$ can be evaluated using the semiclassical WKB approximation (see Eq. (4)) in the case of collision between the $\alpha$ particle and stiff magic or near-magic spherical nuclei at collision energies $E$ below and slightly above barrier. The $\alpha$-nucleus potential is given by Eqs. (5)-(9). The transmission coefficient is approximated by an expression for a parabolic barrier at collision energies higher than or equal to the barrier energy.

\section{UMADAC: input data, discussion and results}

We choose the data for 401 $\alpha$-transitions between the ground states of parent and daughter nuclei with known values of both the half-lives from Refs. \cite{nudat,dk-adndt} and the spins of parent and daughter nuclei from Ref. \cite{ame2012,wallet}. We select the ground-state-to-ground-state transitions, which have a difference between measured \cite{nudat} and evaluated using the atomic mass data \cite{ame2012} $\alpha$-particle energies less than or equal to 10 keV. The recoil energy of the daughter nucleus is taken into account at the evaluation of the $\alpha$-particle energy. The energies of excitation states in parent and daughter nuclei are larger than 10 keV in most of the selected nuclei \cite{nudat}, therefore we choose the true  ground-state-to-ground-state $\alpha$-transitions. Moreover, the experimental data with precisely measured $\alpha$-particle energy are selected by using this limitation. The additional manual selection of transitions is also used for some cases.

Our previous dataset \cite{dk-adndt,dk-emp} contains 344 cases for the ground-state-to-ground-state $\alpha$-transitions. Note some data presented in our old dataset \cite{dk-adndt} (see, for example, the ground-state-to-ground-state $\alpha$-transitions in $^{216,218}$Ra, $^{220}$Th and many other cases) are skipped in the recent data compilation \cite{nudat}. The reason of such data omission in the recent compilation is not known for us. Therefore we make the current dataset for $T_{1/2}$ for the ground-state-to-ground-state $\alpha$-transitions by adding some data from Ref. \cite{dk-adndt}.

The data for $\alpha$-capture cross sections of $^{40}$Ca, $^{44}$Ca, $^{59}$Co, $^{208}$Pb and $^{209}$Bi were taken from Refs. \cite{subfus-exp-ca1,subfus-exp-ca2,subfus-exp-co,subfus-exp-pb}. These data are the same as before \cite{dk-adndt}.

The energy released in $\alpha$-decay between the ground states of parent and daughter nuclei is calculated using recent evaluation of the atomic mass data \cite{ame2012}. The atomic mass consists of nucleons and electrons contributions. The contribution of the binding energy of atomic electrons in the energy of $\alpha$-decay should be separated, because the $\alpha$-decay is the nuclear process. Therefore the released energy of $\alpha$-decay ($Q$), emitted at transition between ground states of the parent and daughter nuclei, is \cite{dk-adndt}
\begin{eqnarray}
Q&=&\Delta M_\mathrm{p} -(\Delta M_\mathrm{d}+\Delta M_\mathrm{\alpha}) \nonumber \\
&+& B_e(Z_\mathrm{p})-[B_e(Z_\mathrm{p})-B_e(Z_\mathrm{\alpha})] ,
\end{eqnarray}
where $\Delta M_\mathrm{p}$, $\Delta M_\mathrm{d}$ and $\Delta M_\mathrm{\alpha}$ are, correspondingly, the mass-excess of parent, daughter, and $\alpha$ nuclei. $B_e(Z)$ is the electron binding energy of the atom with the number of electrons $Z$. The value of the electron binding energy is evaluated by using the high-precision formula \cite{lpt}
\begin{eqnarray}
B_e(Z)=14.4381 \; Z^{2.39}+1.55468 \times 10^{-6} \; Z^{5.35} \; \mathrm{eV}.
\end{eqnarray}

The new Hartree-Fock-Bogoliubov nuclear mass model based on standard forms of Skyrme and pairing functionals was presented recently \cite{hfb27}. This mass model is characterized by a model standard deviation $\sigma = 0.500$ MeV with respect to essentially all the 2353 available mass data for nuclei with neutron and proton numbers larger than 8. The values of deformation parameters $\beta_{2}$ and $\beta_{4}$ obtained in the framework of this accurate model \cite{hfb27} are used in our calculation for specifying the deformation of daughter nuclei.

The $\alpha$-particle emission from nuclei obeys the spin-parity selection rule. Let $j_\mathrm{p}, \pi_\mathrm{p}$ and $j_\mathrm{d}, \pi_\mathrm{d}$ be the spin and parity values of the parent and daughter nuclei respectively. The $\alpha$ particle has zero value of spin and positive parity, therefore the minimal value of angular momentum $\ell_{\rm min}$ at the $\alpha$-transition between states with $j_\mathrm{p}, \pi_\mathrm{p}$ and $j_\mathrm{d}, \pi_\mathrm{d}$ is
\begin{equation}
\ell_{\rm min} = \left\{
\begin{array}{llll}
\Delta_j & {\rm for} \; {\rm even} & \Delta_j & {\rm and} \;\; \pi_\mathrm{p} = \pi_\mathrm{d}, \\
\Delta_j+1 & {\rm for} \; {\rm odd} & \Delta_j & {\rm and} \;\; \pi_\mathrm{p} = \pi_\mathrm{d}, \\
\Delta_j & {\rm for} \; {\rm odd} & \Delta_j & {\rm and} \;\; \pi_\mathrm{p} \neq \pi_\mathrm{d} , \\
\Delta_j+1 & {\rm for} \; {\rm even} & \Delta_j & {\rm and} \;\; \pi_\mathrm{p} \neq \pi_\mathrm{d},
\end{array} \right.
\end{equation}
where $\Delta_j=|j_\mathrm{p}-j_\mathrm{d}|$. The values of $j_\mathrm{p}, \pi_\mathrm{p}$ and $j_\mathrm{d}, \pi_\mathrm{d}$ are taken from Ref. \cite{ame2012}.

The parametrizations of the parameters of the $\alpha$-nucleus potential (8) and the $\alpha$-particle frequency of assault on the barrier (3) have the same form as previously \cite{dk-adndt}
\begin{eqnarray}
V(Q)&=&v_1 + \frac{v_2 Z}{A^{1/3}} + v_3 I
+ \frac{v_4 Q}{A^{1/3}}+\frac{v_5 Y_{20}(\theta) \beta_2}{A^{1/6}},
\\
r_\mathrm{m}(\theta) &=&r_1 + R (1+\beta_{2} Y_{20}(\theta) + \beta_{4} Y_{40}(\theta)),
\\
R&=&r_2 A^{1/3} (1 + r_3 /A+ r_4 I),\\
d&=&d_1+d_2A^{1/3}, \\
\nu&=&19+S+\nu_0 Z^{1/2}A^{1/6}+\nu_1 ((-1)^{\ell}-1) \\
&+&\nu_2\frac{Z}{\sqrt{Q}}+\nu_3 I+\nu_4 \beta_{2}+\nu_5 \beta_{4}+\nu_6 \frac{\ell(\ell+1)}{A^{1/6}} . \nonumber
\end{eqnarray}
Here $A$ and $Z$ are the number of nucleons and protons in the nucleus, which is interacting with the $\alpha$-particle, and $I=(A-2Z)/A=(N-Z)/A$. We search the parameters by using the updated datasets. The procedure of parameters search is described in details in Ref. \cite{dk-adndt}. The obtained parameters values are given in Table I. The values of parameter $S$ equal $5.1304$, $4.4898$, $4.6949$ and $4.5162$ for even-even (e-e), even-odd (e-o), odd-even (o-e), and odd-odd (o-o) nuclei, respectively.

\begin{table}
\caption{Parameters of the $\alpha$-nucleus potential and the assault frequency.}
\begin{tabular}{|l|r|}
\hline \hline
$v_1$ (MeV) & $-41.388$ \\
$v_2$ (MeV) & $-4.0971 \times 10^{-2}$\\
$v_3$ (MeV) & $-18.792$\\
$v_4$ & $0.12885 $\\
$v_5$ (MeV) & $5.9788 \times 10^{-2}$ \\
$r_1$ (fm) & $2.3598$ \\
$r_2$ (fm) & $1.04$\\
$r_3$ & $1.3010$ \\
$r_4$ & $-5.4086 \times 10^{-3}$\\
$d_1$ (fm) & $1.0303$ \\
$d_2$ (fm) & $-2.0881$ \\
$\nu_0$ & $-0.17315$\\
$\nu_{1}$ & $0.87737$\\
$\nu_2$ (MeV$^{-1/2}$) & $ -6.3391 \times 10^{-2}$ \\
$\nu_3$ & $4.2867$\\
$\nu_4$ & $-1.1762$ \\
$\nu_5$ & $-2.1315$\\
$\nu_6$ & $3.9382 \times 10^{-2}$\\
\hline \hline
\end{tabular}
\end{table}

Logarithms of the ratios between theoretical $T_{1/2}^{\rm theor}$ and experimental $T_{1/2}^{\rm exp}$ $\alpha$-decay half-lives versus the mass number $A$ of the parent nucleus are presented in Fig. 1. For most nuclei values of $\log_{10}{\left(T_{1/2}^{\rm theor}/T_{1/2}^{\rm exp}\right)}$ are in the range from -1 to +1. That is, the $\alpha$-decay half-lives evaluated in the framework UMADAC agree well with the experimental data. Note that the values of the $\alpha$-decay half-lives are scattered over an extremely wide range from $\approx 10^{-7}$s to $\approx 10^{+27}$s.

The root-mean-square (RMS) error of the decimal logarithm of $\alpha$-decay half-lives is determined as
\begin{eqnarray}
\delta &=& \sqrt{\frac{1}{N-1} \sum_\mathrm{{k=1}}^\mathrm{N} \left[\log_{10}(T_{1/2}^{\rm theor})-\log_{10}(T_{1/2}^{\rm exp})\right]^2} \\
&=& \sqrt{\frac{1}{N-1} \sum_\mathrm{{k=1}}^\mathrm{N} \left[\log_{10}\left(\frac{T_{1/2}^{\rm theor}}{T_{1/2}^{\rm exp}} \right) \right]^2}. \nonumber
\end{eqnarray}
We use this expression for evaluation of the total $\delta_\mathrm{tot}$ and partial (even-even $\delta_\mathrm{e-e}$, even-odd $\delta_\mathrm{e-o}$, odd-even $\delta_\mathrm{o-e}$ and odd-odd $\delta_\mathrm{o-o}$) RMS errors in the framework of our and other models by using our dataset for $T_{1/2}^{\rm exp}$. The values of RMS errors $\delta_\mathrm{tot}$, $\delta_\mathrm{e-e}$, $\delta_\mathrm{e-o}$, $\delta_\mathrm{o-e}$ and $\delta_\mathrm{o-o}$ obtained in our model are presented in Table II. We see in Table II that the values of these errors are small.

\begin{figure*}
\begin{center}
\includegraphics[width=17.5cm]{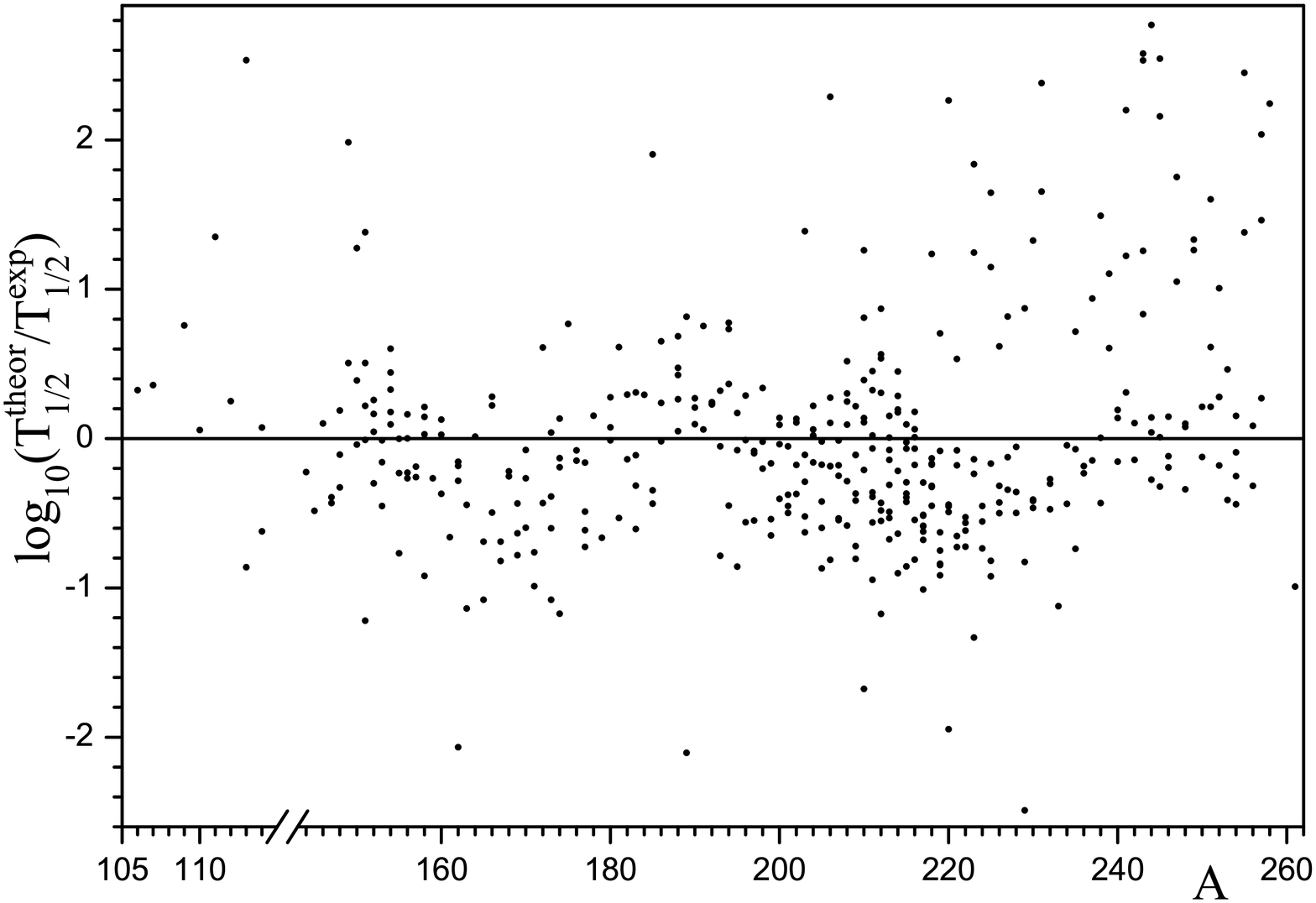}
\caption{(Color online) Logarithms of the ratios between theoretical $T_{1/2}^{\rm theor}$ and experimental $T_{1/2}^{\rm exp}$ $\alpha$-decay half-lives versus the mass number $A$ of the parent nucleus.}
\end{center}
\end{figure*}

\begin{table}
\caption {The RMS errors $\delta$ of the decimal logarithm of the ground-state-to-ground-state $\alpha$-transition half-lives calculated for the full dataset as well as for e-e, e-o, o-e, o-o subsets of the full data set. $N$ is the number of nuclei in the corresponding set or subset. The last column contains the references for corresponding approaches. }
\begin{tabular}{|cccccc|}
\hline \hline
$\delta_{tot}$ & $\delta_{e-e}$ & $\delta_{e-o}$ & $\delta_{o-e} $ & $\delta_{o-o}$ & model \\
($N$=401) & ($N$=145) & ($N$=107) & ($N$=86) & ($N$=63) & \\
\hline
0.590&0.327&0.683&0.696&0.732& SET I \\
0.630&0.393&0.647&0.772&0.821&\cite{royer2011}\\
0.753&0.325&0.956&0.837&0.944& UMADAC \\
1.003&0.404&1.231&1.148&1.311& \cite{wnlg}\\
1.117&0.508&1.232&1.471&1.376&\cite{dasgupta}\\
1.229&0.399&1.529&1.478&1.542&\cite{royer}\\
1.455&1.283&1.452&1.707&1.496&\cite{mnk}\\
1.552&0.714&1.873&1.854&1.909& \cite{pgg}\\
\hline \hline
\end{tabular}
\end{table}

The $\alpha$-capture cross-sections of $^{40}$Ca, $^{44}$Ca, $^{59}$Co, $^{208}$Pb and $^{209}$Bi evaluated in UMADAC agree well with experimental data \cite{subfus-exp-ca1,subfus-exp-ca2,subfus-exp-co,subfus-exp-pb}. The new figure is very similar to Fig. 2 in Ref. \cite{dk-adndt}, therefore we have not presented it.

\section{Empirical relationships for $\alpha$-decay half-lives}

The empirical equation for evaluatiing $\alpha$-decay half-lives in nuclei proposed in Ref. \cite{dk-emp} is written as
\begin{eqnarray}
\log_{10}(T_{1/2})&=&-a - b(A_\mathrm{p}-4)^{1/6} Z_\mathrm{p}^{1/2} +c \frac{Z_\mathrm{p}}{\sqrt{Q}}\\
&+& d \frac{A_\mathrm{p}^{1/6}\sqrt{\ell(\ell+1)}}{Q}- e [(-1)^{\ell}-1]. \nonumber
\end{eqnarray}
Here $A_\mathrm{p}$ and $Z_\mathrm{p}$ are the mass number and charge of the parent nucleus, respectively, and $\ell$ is the orbital moment of the emitted $\alpha$-particle. The value of $T_{1/2}$ is given in seconds; the reaction energy $Q$ and $\ell=\ell_{\rm min}$ are defined by Eqs. (11) and (13), respectively. The coefficients $a,b,c,d,e$ will be specified below.

The first and third terms of the empirical equations were first discussed in Ref. \cite{gn} and after that have been used in various phenomenological expressions for  $\alpha$-decay half-lives. The second term is similar to the corresponding one in Ref. \cite{royer}, but we have introduced additional dependence related to the reduced mass [$\mu \propto 4(A_\mathrm{p}-4)/A_\mathrm{p}$] into this term \cite{dk-emp}, because the $\alpha$-decay half-life evaluated in the WKB approximation depends on the reduced mass, see Eqs. (4) and (9). The second term in Eq. (20) is written after algebraic transformation of the corresponding term in Ref. \cite{dk-emp}. The centrifugal potential (9) distinctly contributes to the total $\alpha$-nucleus potential at small distances between the daughter nucleus and the $\alpha$-particle at $\ell \neq 0$. As a results, the accurate consideration of the $\alpha$-transitions should take into account the spins and parities of parent and daughter nuclei and angular momentum of the emitted $\alpha$-particle, therefore the fourth and fifth terms are added, see Eq. (20).

\subsection{Relationships for full datasets}

We introduce the empirical relationships for $\alpha$-decay half-lives in e-e, e-o, o-e and o-o nuclei. The values of parameters $a,b,c,d,e$ obtained for the full set of data (SET I) are presented in Table III.

\begin{table}
\caption{Values of parameters $a,b,c,d,e$ obtained for various data sets.}
\begin{tabular}{|c|ccccc|}
\hline\hline
Subset & $a$ & $b$ & $c$ & $d$ & $e$ \\
\hline
 & \multicolumn{5}{c|}{Full dataset (SET I)} \\
e-e &26.121&1.1555&1.6085& $-$ & $-$ \\
e-o &30.132&1.0906&1.6981&0.96337&0.60817\\
o-e &29.017&1.1057&1.6745&0.44456&0.63908\\
o-o &32.131&1.0390&1.7370&0.36403&0.56741\\
\hline
 & \multicolumn{5}{c|}{Light nuclei dataset (SET II)} \\
e-e &29.340&1.0380&1.6348& $-$ & $-$ \\
e-o &27.730&1.1079&1.6360&0.68279&0.46841\\
o-e &24.797&1.1894&1.6000&0.46467&0.69250\\
o-o &31.741&1.1052&1.7668&0.28076&0.68208\\
\hline
 & \multicolumn{5}{c|}{Heavy nuclei dataset (SET III)} \\
e-e &28.385&1.0313&1.5847& $-$ & $-$ \\
e-o &31.924&1.0505&1.7226&1.1097&0.48490\\
o-e &36.749&0.79692&1.6863&0.59811&0.51889\\
o-o &40.423&0.48333&1.5976&0.80243&0.27284\\
\hline\hline
\end{tabular}
\end{table}

The RMS errors of the decimal logarithm of the $\alpha$-decay half-lives (19) are presented in Table II. Similar RMS errors obtained by using empirical relationships from Refs. \cite{royer2011,wnlg,dasgupta,royer,mnk,pgg} for our dataset are also given in Table II. We use the same values of parameters in the formulas from Refs. \cite{royer2011,wnlg,dasgupta,royer,mnk,pgg} as recommended in the cited papers. Only the $\alpha$-decay energy is evaluated according to Eq. (11). We stress that we use 401 half-lives for the ground-state-to-ground-state $\alpha$-transitions for evaluating our empirical formulas. In contrast to this some other empirical relationships are obtained by using data for the total half-lives. The universal decay law \cite{pgg} is determined by fitting the half-lives in both $\alpha$-decay and cluster emitters. Our empirical relationships (SET I) have the lowest values of the RMS errors for the total dataset  compared to other empirical relationships. The values of the RMS errors obtained in UMADAC are also very small compared to other approaches.

We introduce into the empirical relationships for $\alpha$-decay half-lives in e-o, o-e, and o-o nuclei two $\ell$-dependent terms, which relate to angular momentum and parity corrections. Note that the empirical relationships obtained in Refs. \cite{dasgupta,royer,mnk,pgg} have not taken into account any angular momentum and/or parity corrections. Due to this the RMS errors obtained by using UMADAC and Eq. (20) with parameters of SET I given in Table III for $\alpha$-transitions in e-o, o-e and o-o nuclei are much smaller than the ones evaluated by using relationships from Refs. \cite{dasgupta,royer,mnk,pgg}, see Table II. The empirical relationships obtained in Refs. \cite{royer2011,wnlg} also included the angular momentum and/or parity correction terms, but the shapes of the angular momentum terms in these Refs. differ from the one presented in Eq. (20). As a result, the RMS errors obtained by using the empirical relationships obtained in Refs. \cite{royer2011,wnlg} are also small. So, the angular momentum and/or parity correction terms of the empirical relationships are very important for accurate description of the $\alpha$-decay half-lives in e-o, o-e, and o-o nuclei. The forms of angular momentum and/or parity correction terms proposed in Eq. (20) and in Ref. \cite{royer2011} lead to smaller values of the RMS errors than the ones in Ref. \cite{wnlg}.

\subsection{Relationships for light and heavy nuclei subsets}

Similar to our previous work \cite{dk-emp} we also apply the empirical equation (20) for evaluation of $\alpha$-decay half-lives for heavy (with $A-Z>126$ and $Z>82$; SET III) and light (rest of the nuclei from the full dataset; SET II) nuclei. The fitting procedure and all definitions are the same as before. As a result, we find coefficients $a,b,c,d,e$ for fitting $T_{1/2}$ in light and heavy subsets of nuclei, which are presented in Table III.

Using our empirical relationships for light and heavy subsets of nuclei we evaluate the RMS errors of $\alpha$-decay half-lives for light and heavy data subsets, see Tables IV and V, respectively. Similar RMS errors obtained with the help of empirical relationships from Refs. \cite{dasgupta,royer,royer2011,mnk,wnlg} and UMADAC for these data subsets are also presented in Tables IV and V. Comparing the results in Tables IV and V we conclude that the empirical relationships (20) with parameters from Table III describe well the $\alpha$-decay half-lives values in the dedicated subset of nuclei.

\begin{table}
\caption{RMS errors $\delta$ of the decimal logarithm of the ground-state-to-ground-state $\alpha$-transition half-lives calculated for the total light data subset as well as for e-e, e-o, o-e, o-o subsets of the total light data subset. The notations are the same as in Table II.}
\begin{tabular}{cccccc}
\hline
\hline
$\delta_{tot}$ & $\delta_{e-e}$ & $\delta_{e-o}$ & $\delta_{o-e} $ & $\delta_{o-o}$ & model \\ ($N=$237) & ($N=$86) & ($N=$65)
& ($N=$46) & ($N=$40) & \\
\hline
0.479&0.285&0.555&0.549&0.601& SET II \\
0.532&0.376&0.596&0.614&0.621& SET I \\
0.601&0.328&0.711&0.681&0.763& UMADAC \\
0.614&0.490&0.560&0.765&0.751& \cite{royer2011} \\
0.733&0.490&0.586&0.913&1.098& \cite{wnlg} \\
0.869&0.493&0.724&1.290&1.121& \cite{royer} \\
0.875&0.585&0.652&1.423&0.926& \cite{dasgupta} \\
1.498&0.653&1.030&1.889&1.326& \cite{mnk} \\
\hline\hline
\end{tabular}
\end{table}

\begin{table}
\caption{The RMS errors $\delta$ of the decimal logarithm of the ground-state-to-ground-state $\alpha$-transition half-lives calculated for the total heavy data subset as well as for e-e, e-o, o-e, o-o subsets of the total heavy data subset. The notations are the same as in Table II.}
\begin{tabular}{cccccc}
\hline
\hline
$\delta_{tot}$ & $\delta_{e-e}$ & $\delta_{e-o}$ & $\delta_{o-e} $ & $\delta_{o-o}$ & model  \\
 ($N=$164) & ($N=$59) & ($N=$42)
& ($N=$40) & ($N=$23) & \\
\hline
0.612&0.138&0.786&0.714&0.816& SET III \\
0.657&0.175&0.770&0.790&0.950& \cite{royer2011} \\
0.667&0.241&0.806&0.788&0.909& SET I \\
0.931&0.301&1.255&0.996&1.218& UMADAC \\
1.300&0.231&1.838&1.381&1.645& \cite{wnlg} \\
1.396&0.280&1.948&1.495&1.788& \cite{mnk} \\
1.398&0.375&1.805&1.543&1.954& \cite{dasgupta} \\
1.616&0.196&2.286&1.686&2.115& \cite{royer} \\
\hline
\hline
\end{tabular}
\end{table}

\section{Conclusion}

We found the parameter values of UMADAC. These values have been obtained by using the updated values of the $\alpha$-decay half-lives, the binding energies of nuclei, the spins of parent and daughter nuclei, and the surface deformation parameters of daughter nuclei. The data for $\alpha$-decay half-lives in spherical and deformed nuclei and for $\alpha$-capture reactions are well described in the framework of UMADAC.

We determined the updated parameter values of simple empirical relationship for $\alpha$-decay half-lives for the total, light, and heavy subsets of nuclei. The available $\alpha$-decay half-lives values are well described by the empirical relationship for $\alpha$-decay half-lives which take into account the spin-parity properties of the parent and daughter nuclei.

\end{document}